\documentclass[epjST]{svjour}
\usepackage{graphicx}        % standard LaTeX graphics tool
\usepackage{amsmath}
\usepackage{txfonts}
\usepackage{epsfig}
\usepackage{graphics}
\usepackage{floatrow}
\usepackage[toc,page]{appendix}

\newcommand{\vivo}{\textit{in vivo }}
\newcommand{\vitro}{\textit{in vitro }}

\newcommand{\Var}{\mathbb V\mathrm{ar}}

\newfloatcommand{capbtabbox}{table}[][\FBwidth]

\begin{document}
\title{Fluctuation effects in bidirectional cargo transport.}
\author{Sarah Klein\inst{1,2} \and C\'ecile Appert-Rolland\inst{1} \and Ludger Santen\inst{2}}

\institute{                    
  \inst{1}Laboratory of Theoretical Physics, CNRS (UMR 8627), University Paris-Sud
B\^atiment 210, F-91405 ORSAY Cedex, France\\
  \inst{2} Fachrichtung Theoretische Physik, Universit\"at des Saarlandes D-66123 Saarbr\"ucken, Germany}

\abstract{
We discuss a theoretical model for bidirectional cargo transport in biological cells, which is driven by teams of molecular motors and subject to thermal fluctuations. The model describes explicitly the directed motion of the molecular motors on the filament. The motor-cargo coupling is implemented via linear springs. By means of extensive Monte Carlo simulations we show that the model describes the experimentally observed regimes of anomalous diffusion, i.e. subdiffusive behavior at short times followed by superdiffusion at intermediate times. The model results indicate that subdiffuse regime is induced by thermal fluctuations while the superdiffusive motion is generated by correlations of the motors' activity. We also tested the efficiency of bidirectional cargo transport in crowded areas by measuring its ability to pass barriers with increased viscosity. Our results show a remarkable gain of efficiency for high viscosities.
}

\maketitle

\section{Introduction}
{Many cellular functions depend on active transport processes, which are driven by molecular motors. Molecular motors are proteins which are able to 
perform directed motion along the intracellular network of biopolymers, i.e. the cytoskeleton. The cytoskeleton consists of three different kinds of filaments - microtubules, actin-filaments and intermediate filaments. Transport on large length scales occurs mainly along microtubules from the cell center to the membrane and \textit{vice versa} \cite{alberts2002}. Microtubules are long, polarized biopolymers with a well-defined plus- and minus-end. Molecular motors, in the case of microtubules the families of kinesins and dyneins, can perform steps preferentially to the plus-end (kinesin) or to the minus-end, respectively, under the consumption of Adenosin triphosphate (ATP). In recent years much knowledge has been accumulated in particular with respect to the properties of single molecular motors \cite{Carter,vale2000}. Despite this progress many rather fundamental questions of motor-driven transport are still not answered. This is particularly true for systems of interacting
  motor proteins, as for example a cargo that is driven by teams of molecular motors. Depending on the configuration of the attached motors cargos can be transported uni- or bidirectionally \cite{welte1998,soppina2009}.
Actually,} for several kinds of cargos - for example for endosomes \cite{soppina2009}, mitochondria \cite{hollenbeck_s2005} or lipid droplets \cite{welte1998} - it was observed that they move in a saltatory manner: the trajectories show pieces with persistent motion in one direction and then sudden turns in the other direction. {These properties of the cargo trajectories evidence that both types of motors can apply forces on the cargo.} Besides, the motor-cargo complex shows interesting statistical properties which has been characterized as anomalous diffusive behavior. The mainly disputed point is whether a coordination mechanism is needed to control the interplay between the two teams of motors \cite{welte2004} or if stochastic fluctuations are sufficient to produce this kind of cargo motion \cite{mueller_k_l2008}. We explore in this work whether the second scenario can explain the observed characteristics of cargo trajectories.

In experiments the cargo motion is often characterized via the mean square displacement (MSD) $ \langle (X(t + \Delta t) - X(t))^2\rangle$ of cargo trajectories $X(t)$. The brackets here indicate the average over $t$. For a ballistic motion the MSD is proportional to $t^2$ while it is linear in $t$ for the purely diffusive case without bias. In several \vivo experiments \cite{kulic2008,caspi_g_e2002,salman2002}  it was detected that the cargo's MSD shows a time-dependence $\Delta t^\gamma$ with exponents $\gamma<1$ and $1<\gamma<2$, depending on the time scale ({anomalous diffusion}). In fact, the {time dependence of the MSD is difficult to interpret for finite times. Indeed,  apparent superdiffusion ($\gamma >1$) may originate either from a biased but uncorrelated motion of the cargo or indicate positive temporal correlations of the cargo's displacements. In the latter case it is, compared to normal diffusion, more likely that the cargo continues its motion in the same direction.} 
{The difference between these two kinds of particle motion can be easily distinguished by analyzing the variance
\begin{align}\label{vari}
\Var[X(\Delta t)] = \langle (X(t + \Delta t) - X(t))^2 \rangle - \langle(X(t + \Delta t) - X(t))\rangle^2.
\end{align} 
 instead of the MSD.} We showed in a preceding publication \cite{EPL14} that a cargo transported by two teams of motors with different single motor properties most time exhibits a biased motion, so that variance and MSD are not equal. We shall focus on the variance throughout this paper.

It was found for \vivo experiments \cite{kulic2008,salman2002} that for small time lags the cargo moves mainly subdiffusively and crosses over to a superdiffusive motion for intermediate time lags. On long time scales, way bigger than the average turning time, the motion becomes (sub)diffusive again \cite{caspi_g_e2002}. It was suggested that the cargo {exhibits anomalous diffusive behavior} because of the inner cellular structure that presents several obstacles
which can impede cargo's motion \cite{weiss2004,caspi_g_e2002}. Here we want to study the statistics of cargo trajectories in the absence of such additional effects due to the inner cell structure.

Another explanation of the observed anomalous diffusion could be the motion of microtubules which bend and than relax again and so add a velocity component to the cargo motion \cite{kulic2008}.  
Also the underlying network can influence the MSD or variance. On a branched network a purely ballistic motion also shows $\Var[X] \sim \Delta t^\gamma$ with $1<\gamma<2$ depending on the turning angle distribution \cite{reza2014}. In this work, as we want to ignore all network effects we propagate the cargo along a static one dimensional track.

The model introduced in this article describes bidirectional cargo transport mediated by two different teams of molecular motors. We will show that a subdiffusive ($\Var[x_C(t)] \sim \Delta t^\gamma$ with $\gamma<1$) motion occurs at small time scales if the thermal fluctuations of the cellular environment are taken into account. Besides, superdiffusive ($1<\gamma<2$) motion occurs at longer time scales and no further interaction with the environment is needed to observe this anomalous diffusion.
In a second part, we analyze how the motion is influenced by a viscous barrier in the system representing highly crowded areas. {Having two teams of motors attached to the cargo provides in this scenario an efficient mechanism to pass highly crowded areas.}

\section{Model}
\begin{figure}[tb]
\includegraphics[width = 0.5\textwidth]{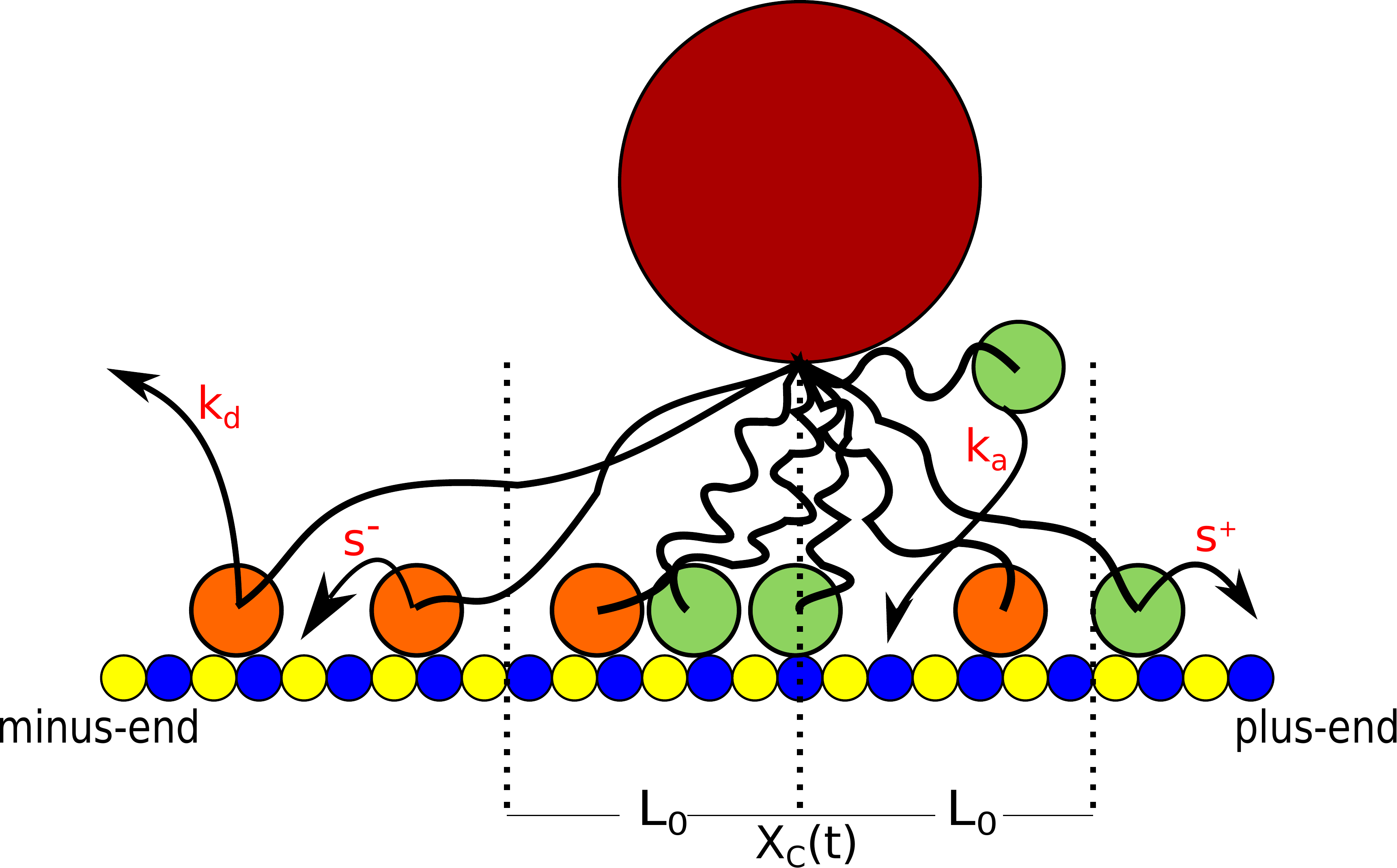}
\caption{Schematic drawing of the motor kinetics. A cargo (red) is moved by two teams of motors pulling in "+" (light green) and "-" (orange) direction, respectively. The single motors can walk on and detach from the filament. Once they are detached, they can attach again within the force free area $[x_C(t)- L_0,x_C(t) + L_0]$.}\label{skizze}
\end{figure}

In this work we introduce a model to describe bidirectional cargo transport by teams of molecular motors. Each motor consists of a head which is bound to the filament at position $x_i$ and a tail, the so-called neck linker, connecting the head and the cargo. To calculate the force $F_i$ produced by these motors and applied on the cargo we take the position of each single motor head into account as shown in Fig. \ref{skizze}.  {This kind of model for motor-cargo complexes has already been }introduced in \cite{korn2009,kunwar2008} and \cite{kunwar2011} with one and two teams of motors, respectively. Since we want to compare our simulation results with \vivo experiments, we model the differences between the two different kinds of motors in detail.

We analyze the motion of a cargo at position $x_C(t)$ at time $t$ pulled by two teams of molecular motors, each consisting of $N=5$ motors. We assume that the neck linker of motors can be modeled as a linear spring with spring constant $\alpha$ and an untensioned length $L_0$ such that the motors exerts no force on the cargo as long as $|x_C(t) - x_i|<L_0$. The motors are tightly bound to the cargo but can detach with a force dependent rate $k_d^\pm(F_i)$ from the filament, where the superscript $\pm$ gives the rate for $+$ and $-$ motors, respectively. Once detached from the filament, the motors can {reattach to the filament  with a constant rate $k_a$. The motors attach within a region $x_C(t)\pm L_0$ and therefore apply no force.} We introduce this untensioned length because a motor that binds to the filament will not directly apply a force, since the motor's neck is not spontaneously stretched but 
%will become so 
by the motion of the motor's head along the filament. The total force applied on the cargo by the $n_+ / n_-$ puling $+/-$-motors then reads
\begin{flalign}
&F (x_C(t),\{x_i\}) =  \sum_{i=1}^{n_++n_-}  F_i (x_i-x_C(t))  \nonumber \\ 
&\text{with   }  F_i (x_i -x_C(t)) = 
\begin{cases}
\alpha (x_i-x_C(t) +L_0), \ \ \  \ &  x_i-x_C(t)<-L_0\\
0 , & |x_i-x_C(t)|<L_0\\
\alpha (x_i-x_C(t)-L_0),  &x_i-x_C(t)>L_0
\end{cases} \nonumber
\end{flalign}
As long as the motor is bound to the filament it can perform steps with the force dependent rate $s^\pm(F_i)$.
The model's dynamics are schematically represented in Fig. \ref{skizze}.

As already mentioned in the introduction{ the motor properties of the two teams are different if realistic biological parameters are used}. That is why we use for stepping \cite{schnitzer_v_b2000,toba2006} and detachment \cite{kunwar2011} rates some expressions based on experimental results and also take the influence of ATP concentration into account. The detachment rate for plus motors (kinesin) is given by
\begin{align}\label{kinesin}
k_d^+(F_i) = \begin{cases}
k_d^0 \exp\left(\frac{|F_i|}{2.5f}\right) &F_i<F_S \\
k_d^0 \left(0.186\frac{|F_i|}{f} + 1.535\right) &F_i \geq F_S 
\end{cases}
\end{align} 
and for minus motors (dynein) by
\begin{equation}\label{dynein}
k_d^-(F_i) = \begin{cases}
k_d^0 \exp\left(\frac{|F_i|}{2.5f}\right) \! &F_i>-F_S \\
k_d^0 \left[1.5\left(1-\exp\left(\frac{-|F_i|}{1.97f}\right)\right)\right]^{-1} \! &F_i\leq -F_S 
\end{cases}.
\end{equation}
with the force-free detachment rate $k_d^0$ \cite{kunwar2011}%NEW
\footnote{Since the \vivo and \vitro behavior differs significantly in \cite{kunwar2011} some model-parameters in the 
	detachment rate expressions have been adjusted in order to avoid dominance of one motor species. 
	 Here we chose slightly different parameter values to achieve the same goal.}. 
	 In \eqref{kinesin} and \eqref{dynein} we introduce two force scales - the stall force $F_S$ which is the maximum force under which the stepping rate in the preferred motor direction does not vanish and a normalization force $f = 1 $ pN to get the right units. 

For the stepping rate $s^\pm(|F_i|,[ATP])$ in the region of forces smaller than the stall force we use a two-state Michaelis-Menten equation as suggested in \cite{schnitzer_v_b2000}
\begin{equation}\label{MMe}
s (|F_i|,[ATP])= \frac{k_\text{cat}(|F_i|)[ATP]}{[ATP] + {k_\text{cat}(|F_i|)}{k_\text{b}(|F_i|)}^{-1}},
\end{equation}
with the catalytic-turnover rate constant $k_\text{cat}(|F_i|)$ and a second-order rate constant for
ATP binding $k_\text{b}(|F_i|)$. Schnitzer \textit{et al.} \cite{schnitzer_v_b2000} also introduce a Boltzmann-type force relation for the rate constants
\begin{equation}\label{delta}
k_j(F_i) = \frac{k_j^0}{p_j + q_j \exp(\beta F_i\Delta)} \ \ \ \ \ \ j=\{\text{cat, b} \}
\end{equation} 
with constants $k_j$, $p_j + q_j = 1$, $\beta = (k_\text{b} T)^{-1}$ and $\Delta$ (see \cite{schnitzer_v_b2000} for more details). It was measured for kinesin \cite{schnitzer_v_b2000} and dynein \cite{toba2006} that the stepping rate, depending on [ATP] and the load force $F_i$, can be described by eq. (\ref{MMe}). If the force on a motor is bigger than the stall force the motors step backwards with a constant rate $s_b^\pm=v_b/d$.

The stall force for minus motors is taken to vary in a linear affine manner from 0.3 pN at vanishing ATP concentration up to 1.2 pN for saturating ATP levels \cite{mallik_g2004} while we leave kinesin's stall force constant at 2.6 pN \cite{kunwar2011}. This determines $\Delta$ as defined in eq. \eqref{delta} to ensure that the stepping rate is zero at stall. 

Now knowing the motor kinetics we further have to define how a cargo with mass $m$ and radius $R$ reacts on these forces. We describe the motion of the cargo via a Langevin equation
\begin{align}
m \frac{\partial^2 x_C(t) }{\partial t^2 } = -\beta \frac{\partial x_C(t)}{\partial t} + F(x_C(t),\{x_i\}) + F_{therm}(t)
\label{eqofm}
\end{align}
with Stokes' friction $\beta = 6 \pi \eta R$ due to the cytosol's viscosity $\eta$ and the stochastic force $F_{therm}(t) = \sqrt{2k_B T \beta}\xi(t)$ due to thermal noise inside the cell. Here $k_B$ is the Boltzmann constant, $T$ the temperature and $\xi(t)$ a normalized white-noise process, hence
\begin{equation}
\langle \xi(t) \rangle = 0  \ \ \ \text{ and }  \ \ \ \langle \xi(t)\xi(t') \rangle  = \delta(t-t') \ \ \ \forall \ t,t'.
\end{equation}

The cargo has to be propagated in continuous time
according to equation~\eqref{eqofm} between two motors
events (stepping, detachment or attachment).
In constrast to our previous
work~\cite{EPL14}, this evolution equation
contains some thermal
noise which requires a specific treatment.
In~\cite{gillespie1996,norrelykke2011},
a stochastic procedure was proposed in order
to generate at discrete times some fluctuations in
the cargo trajectory,
with the same amplitude as would result from the
integration of the Langevin equation.
We follow this procedure and define the moments
at which these thermal fluctuations are included
as ``shot events''.

At each shot event $E_i$ at time $t_{E_i}$, two independent
random numbers
$\varphi_{i}$ and $\zeta_i$ are chosen from
a zero-mean, unit-variance Gaussian distribution.
Then, some contributions of thermal noise
to the position and
velocity of the cargos are build up until the
next shot event, according to the
expressions~\cite{gillespie1996,norrelykke2011} %NEW
\begin{align}\label{sig}
x_C^{therm}(t-t_{E_i})  =  \sigma_{xx}(t-t_{E_i}) \varphi_i 
\end{align}
\begin{align}\label{sig2}
v_C^{therm}(t-t_{E_i})  =  \frac{\sigma_{xv}(t-t_{E_i})^2}{\sigma_{xx}(t-t_{E_i})}\varphi_i+ \sqrt{\sigma_{vv}(t-t_{E_i})^2 - \frac{\sigma_{xv}(t-t_{E_i})^4}{\sigma_{xx}(t-t_{E_i})^2}} \zeta_{i}.
\end{align}

The expressions for the time-dependent $\sigma_{xx}(t-t_{E_i}),\ \sigma_{xv}(t-t_{E_i})$ and
$\sigma_{vv}(t-t_{E_i})$ are given in Appendix \ref{app} and
verify
$x_C^{therm}(0) = v_C^{therm}(0)  = 0$.
At the next shot event $E_{i+1}$ (we assume here
that no motor event occured in the meantime),
the built-up thermal fluctuations are
added to the cargo components before drawing
new random numbers :
\begin{eqnarray}
x_C(t_{E_{i+1}}) & = & x_C^d(t_{E_{i+1}})
+ x_C^{therm}(t_{E_{i+1}}-t_{E_i})
\label{shot1}\\
v_C(t_{E_{i+1}}) & = & v_C^d(t_{E_{i+1}})
+ v_C^{therm}(t_{E_{i+1}}-t_{E_i})
\label{shot2}
\end{eqnarray}
Here $x_C^d$ and $v_C^d$ are the deterministic cargo
position and velocity calculated by solving the
equation of motion \eqref{eqofm} without the
stochastic force $F_{therm}(t)$ as described in
\cite{TGF13} and with the initial condition at $t_{E_i}$ to be at
position $x_C(t_{E_i})$ with $v_C(t_{E_i})$.

The history
of the system is punctuated by two types of events,
motor events and shot events.
The knowledge of the cargo position at every time
$t$ is necessary to get the force on the cargo
at every arbitrary time so that we can use
Gillespie's %next-event algorithm 
for time-dependent
rates \cite{gillespie1978}, in order to predict
which event will occur next, and when.
In order to know the cargo position
not only at discrete times as in~\eqref{shot1} but
in continuous time, we interpolate between
the shot events by generalizing the
expression~\eqref{shot1} to
\begin{equation}
x_C(t)  =  x_C^d(t) + x_C^{therm}(t-t_{E_i})
\label{interpol}
\end{equation}
for all times $t_{E_i} \leq t < t_{E_j}$
between the two successive shot events $E_i$ and $E_j$.
While the thermal contributions added at discrete
times have the same statistics than those
that would be obtained from the direct
solution of the Langevin equation,
as proved in~\cite{gillespie1996,norrelykke2011},
the interpolation in~\eqref{interpol}
is an approximation, which
should be correct if the shot frequency is
high enough.
Still, as we want to analyze the long time behavior,
the frequency cannot be too high due to 
computational limitations.
We chose to let the thermal shots occur
with a constant rate $k_s$ which is at least
hundred times bigger than the average single motor
event rate, thus in the order of $k_s=10^5$ s$^{-1}$.
We checked that our choice for $k_S$ is sufficient to avoid discretization effects.

For completeness all simulation parameters are given in Table \ref{paraset}. 
\section{Results and Discussion}
In this section we first concentrate on the time-dependent variance of the cargo motion and analyze how it evolves with time. Secondly, we introduce a viscous barrier in the system and investigate how this change in the effective viscosity influences the cargo's transport efficiency.
\subsection{Cargo's displacement Variance}
\begin{figure}[bt]
\centering
\includegraphics[width=0.5\textwidth]{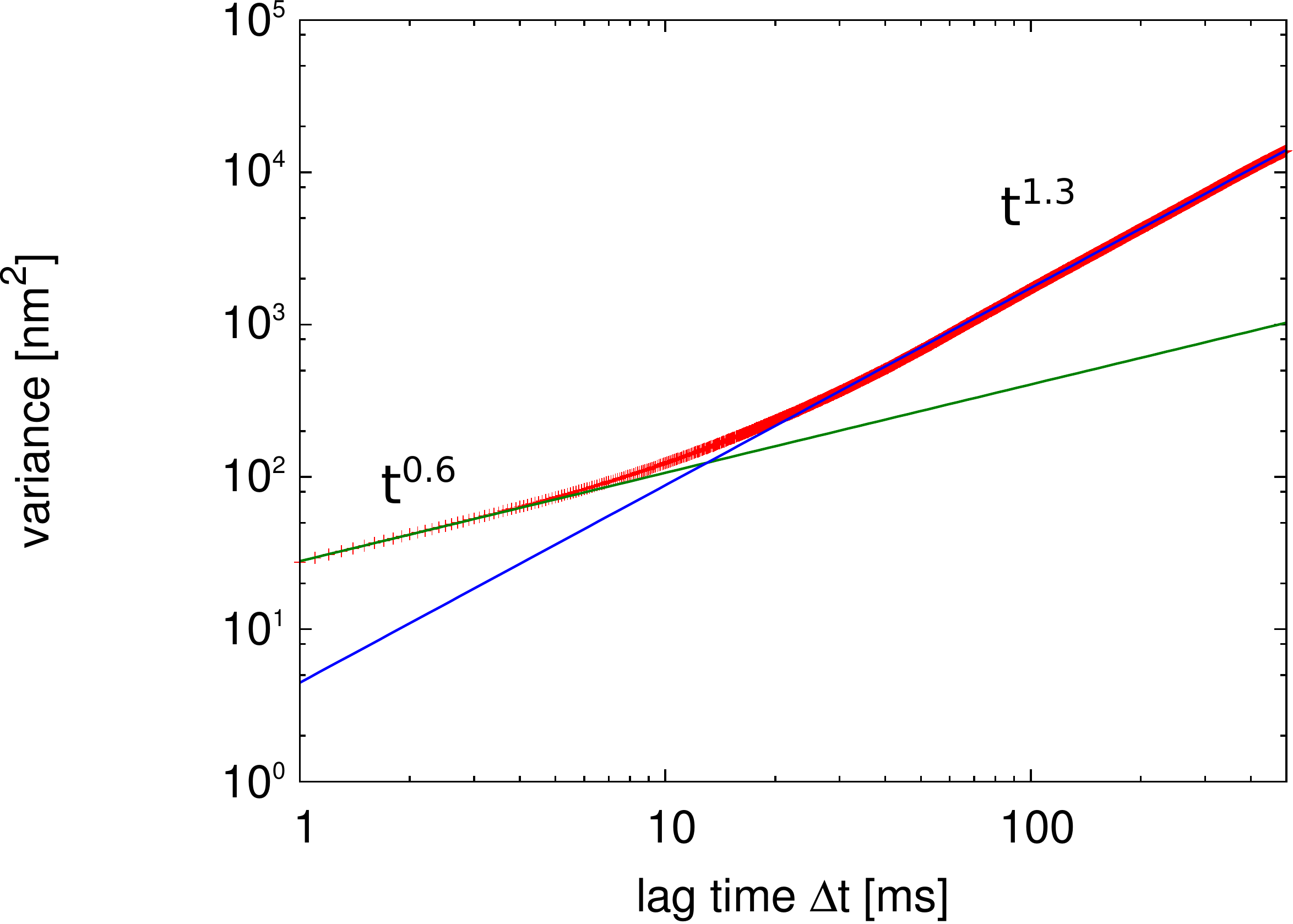}
\caption{Variance of the cargo pulled by $N+=N_-=5$ motors through a viscous medium. At short times the cargo shows subdiffusive behavior $ \sim \Delta t^\gamma$ with an exponent $\gamma=0.6$. For times $t>10$ ms the cargo moves superdiffusively with $\gamma=1.3$.}
\label{msd}
\end{figure}

In this first subsection we calculate the variance of the cargo trajectories as given in eq.~\eqref{vari} {where we carefully checked that the results are independent from the initial conditions.}  Using the set of parameters given in Table \ref{paraset} we observe subdiffusive motion $\Var[x_C(t)] \sim \Delta t^\gamma$ with an exponent $\gamma=0.6$ for times smaller than $10$ ms and superdiffusion for larger times with an exponent $\gamma =1.3$ as shown in Fig. \ref{msd}. {The subdiffusive cargo motion can be attributed to  thermal fluctuations in the external potential} of the motor springs whereas the superdiffusion is observed due to the correlation of the motor stepping events. This is for our chosen set of parameters observable on time scales of several hundred motor steps. We have already shown in \cite{EPL14} that without thermal noise the superdiffusive behavior is observed while subdiffusion is not. 

{Recently, the MSD has been calculated from particle trajectories in Drosophila S2 cells~\cite{kulic2008}. In this work a crossover from sub- to superdiffusive has been reported at $\Delta t \approx 30$ms. In the subdiffusive regime, $\gamma = 0.59 \pm 0.28$ has been obtained, where exponents are varying from 0.2 to 1.2 for different trajectories. At times  $\Delta t > 30$ms, a mean value  $\gamma = 1.62 \pm 0.29$ has been established, where the results for single trajectories are in the range of 1.2 to 2. We notice a remarkable agreement between experimental and model results in the subdiffusive regime (for which we checked that variance and MSD do not differ). The results for $\gamma$ in the superdiffusive regime strongly depend on the bias of the cargo if the MSD is considered and not the variance of the particle position. Taking this into account our results are at least not contradicting the experimental findings. However, in order to test the actual agreement between mo
 del and experimental results the variance should be taken into account which characterizes more directly the correlation of the motors' dynamics. } 

\subsection{Viscous barrier}
In a crowded compartment of a cell the effective viscosity can be enhanced by a factor up to 1000 in comparison to the viscosity of pure water~\cite{luby1999}. In a previous publication we have shown that the above described model exhibits non-monotonous dependence of the bias on the viscosity~\cite{EPL14}. This motivated us to analyze the influence of spatial confinement in a crowded area of the cell on the cargo dynamics. Crowded areas are considered as regions of high effective viscosity. In order to assess the mobility of the motor-cargo complex we compare its motion to pure diffusion in the same environment.

\begin{figure}[bt]
\includegraphics[width=0.5\textwidth]{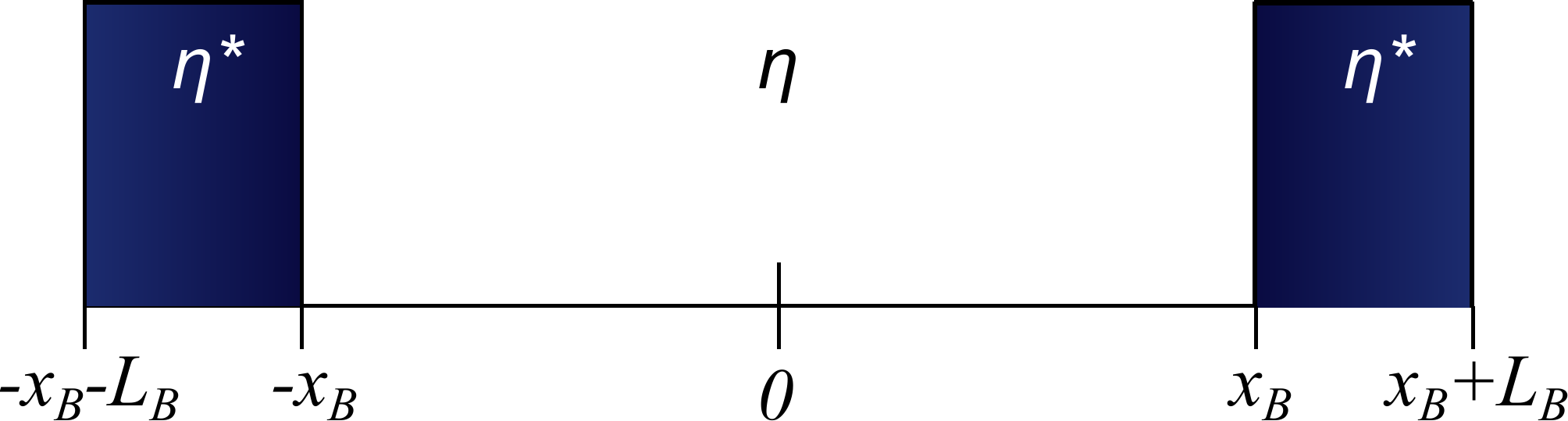}
\caption{Schematic drawing of the arrangement of the barriers. The blue rectangles represent the area of increased viscosity $\eta^*$ .}\label{barr_draw}
\end{figure}

\begin{figure}[htb]
\begin{minipage}{0.5\textwidth}\includegraphics[width=0.6\textwidth,angle=270]{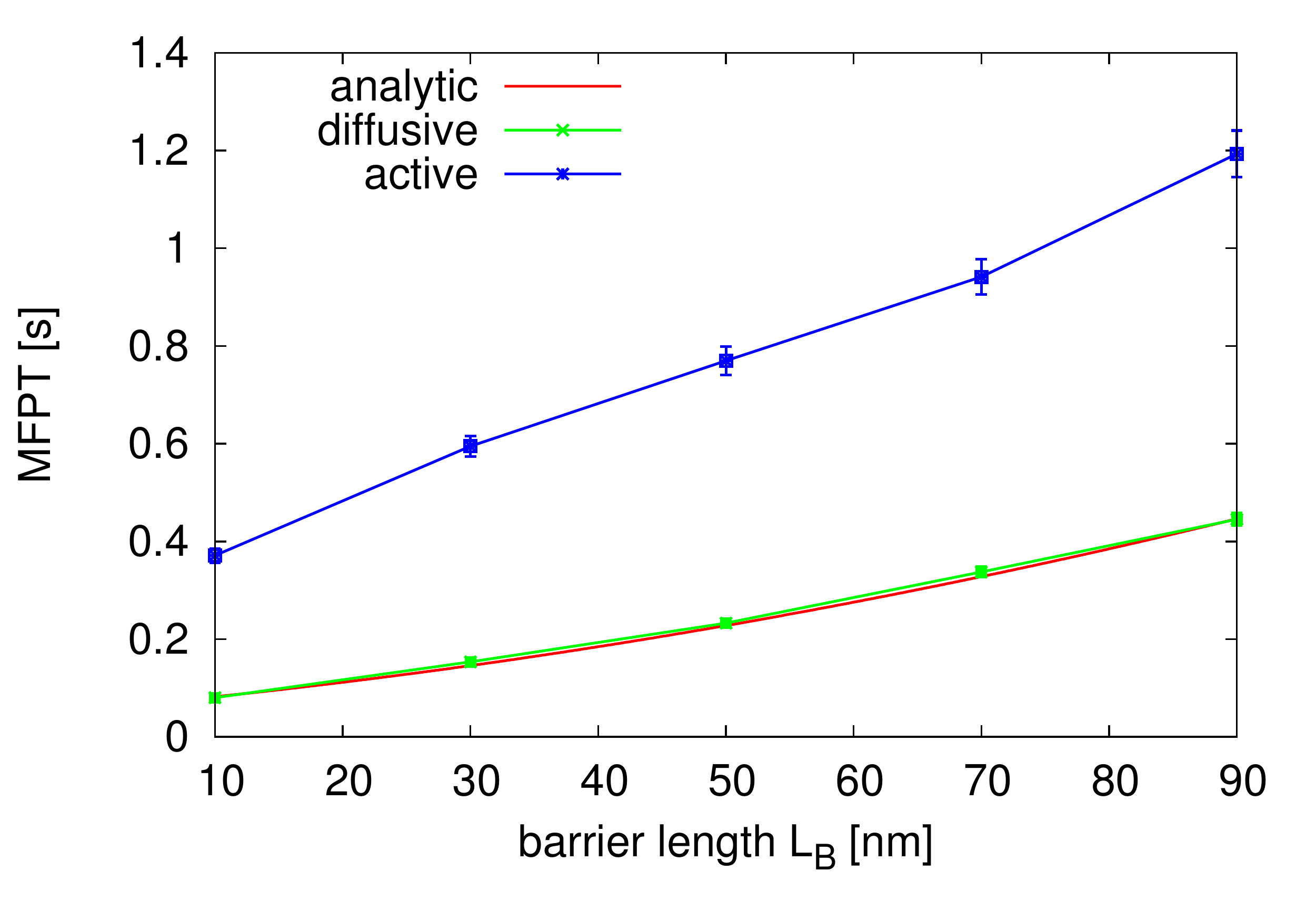}\\ \hspace*{0.5\textwidth}\textbf{(a)}\end{minipage}\begin{minipage}{0.5\textwidth}
\includegraphics[width=0.6\textwidth,angle=270]{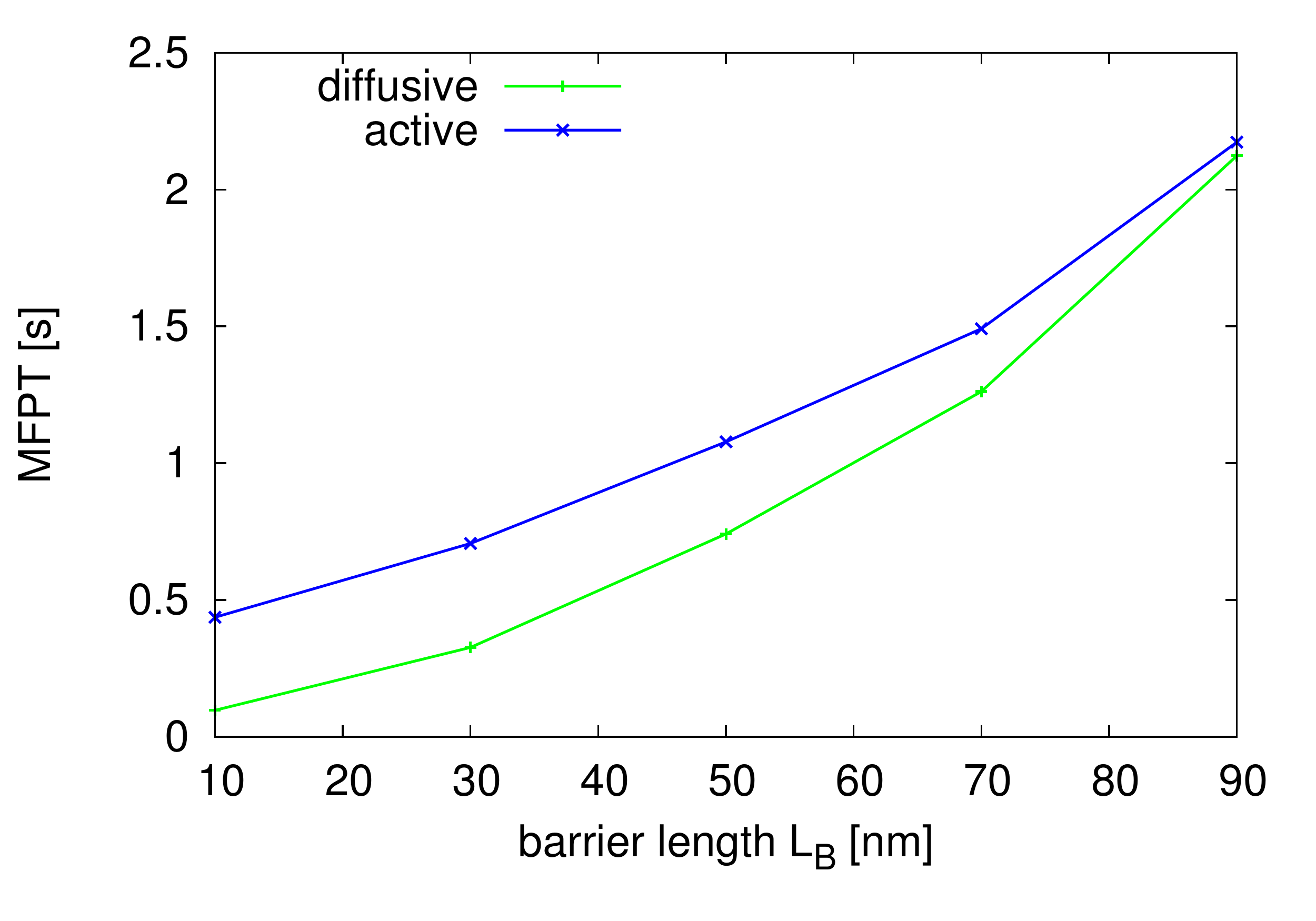}\\ \hspace*{0.5\textwidth}  \textbf{(b)}\end{minipage}\\
\begin{minipage}{0.45\textwidth}
\includegraphics[width=0.7\textwidth,angle=270]{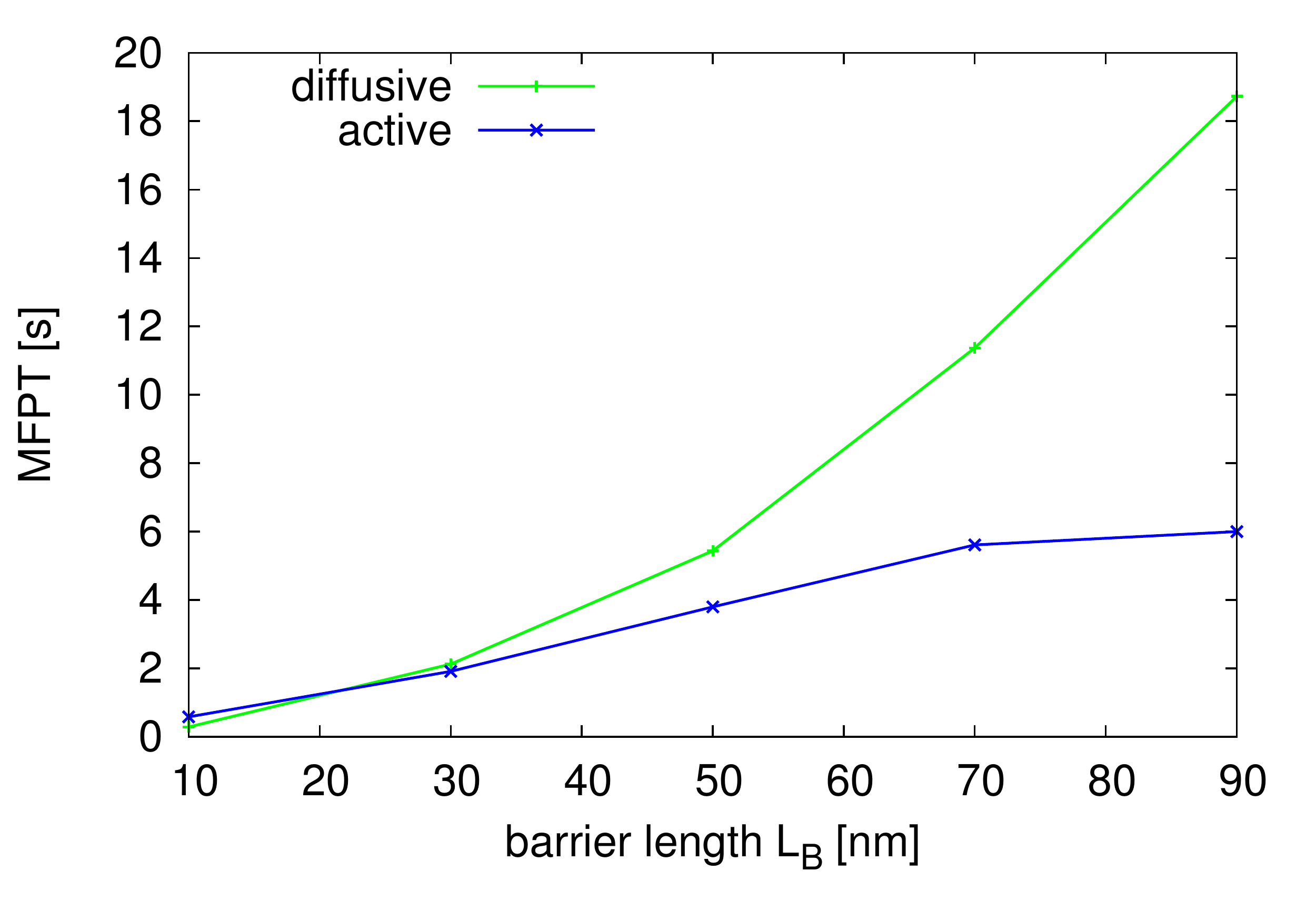}  \hspace*{0.5\textwidth}\textbf{(c)}\end{minipage}\begin{minipage}{0.45\textwidth}\caption{MFPT for different barrier viscosities (\textbf{(a)} $\eta^*=\eta$, \textbf{(b)} $\eta^* = 10 \eta$, \textbf{(c)} $\eta^* = 100\eta$) and different barrier lengths $L_B$. We show the cases with active transport (blue) and pure diffusion (green). In some regimes it is more efficient to diffuse through the cell, especially for small distances and low viscosities. If the effective viscosity is considerably higher it is more productive to actively transport the cargo. The red line in \textbf{(a)} shows the analytic solution for the MFPT for diffusion on an interval. The errorbars give the Gaussian error.}\label{barrier}
\end{minipage}
\end{figure}

We introduce two viscous barriers with increased effective viscosity $\eta^{*}$  and with a given length $L_B$ at positions $\pm x_B$ which represent a highly crowded area (see Fig. \ref{barr_draw}). In Fig. \ref{barrier} the mean first passage time (MFPT) to cross the barrier at position $\pm x_B \pm L_B$ starting at position $x_0 =0$ is shown. To make conclusions about the time needed to cross the barrier we compare it on the one hand to the purely diffusive case (green curve in Fig. \ref{barrier}) and also to the barrier-free case (Fig. \ref{barrier}\textbf{(a)}).

{Interestingly, in the case of low barrier viscosities $\eta^*$ and unbiased cargo dynamics (Fig. \ref{barrier}\textbf{(a)}) pure diffusion outperforms the active transport of cargo. 
Irrespective of the chosen barrier length our results for the MFPT are much longer in case of actively transported cargos as compared to diffusion, for which exact results for the MFPT }
\begin{equation}
 \frac{\beta}{2k_BT}(x_B+L_B)^2
\end{equation}
are known \cite{redner2001}, with $\beta = 6\pi\eta R$. With slightly increased viscosity in the barrier ($\eta^* = 10\eta$) the diffusion is still faster, especially for small barriers but one can already recognize that for $L_B=90$ nm the MFPT for diffusion and active transport are already the same in the range of error. If the effective viscosity is further increased inside the barrier ($\eta^* = 100\eta$) the active transport is significantly faster (\ref{barrier}\textbf{(c)}) and the MFPT does not show the quadratic behavior with the interval length $2\cdot(x_B +L_B)$ anymore.

\section{Conclusion}

{In this article we analyzed a model for bidirectional cargo transport driven
by teams of molecular motors. We studied the impact on the cargo’s motion
of thermal noise, and also of crowded areas of the cell, which we represented by a
spatially structured effective viscosity.}

First,  we analyzed the change in the cargo displacement variance if thermal noise
is taken into account. %NEW
With the given parameter combination in this paper 
we have shown that for times smaller than 10 ms the trajectories exhibit
subdiffusive behavior with $\Var[x_C(t)] \sim \Delta t^{0.6}$ which transfers into
superdiffusive behavior at longer times. {The subdiffusive motion is a result
of the thermal fluctuations of the cargo's position, which is trapped in the external
potential of the motor springs. If this potential is stronger (for example stiffer springs, lower detachment rate, higher attachment rate)
we expect that the subdiffusive exponent will decrease. We actually checked that this is the case for stiffer springs.
Contrary, superdiffusion can be obtained because the cargo 
motion has a finite correlation time and tends to continue moving in the same direction. }

In \cite{kulic2008} it was concluded that the observed anomalous diffusion of cell
organelles 
cannot be described solely in terms of cargo movement along stationary microtubule
tracks, but instead includes a strong contribution from the movement of the tracks.
Our results question this interpretation since in our model the motor-cargo complex moves on a
single and infinite track, i.e. the structure and dynamics of the MT-network has not
at all been taken into account and still we find the same anomalous diffusion. 

Additionally, we characterized the influence of a change in the effective
viscosity representing differently crowded areas. For low
viscosities or small areas of increased viscosity the pure cargo diffusion is faster
than an actively transported one, %NEW
since in the latter case the cargo can be trapped in the potential
of motor springs. In crowded areas of the cell, however, the situation is inverted:
While the diffusively moving cargo slows down completely, the actively transported
one keeps its motility to a large extend. %NEW
This illustrates that the cell can use the asymmetry 
between the motors as a driving force.

In this paper, the effect of the environment was modeled as an effective viscosity. Within our modeling approach it is also possible to treat external forces explicitly. 

It would be interesting to build experiments that would allow direct comparison with the model. Very recently it became possible to use dynein in motility assays \cite{mckenney2014}. Due to this
achievement it will also be possible to study bidirectionally transported
motor-cargo complexes \vitro. Such experiments could be used in order to
validate the modeling approach. Then both, experiment and model, could be modified to account for more complex situations, example giving the influence of well-defined network structures.

\appendix
\setcounter{equation}{0}
\def\@newusecounter#1{\@nmbrlisttrue\def\@listctr{#1}}
\renewcommand{\theequation}{A\arabic{equation}}
\section{Thermal noise propagation}\label{app}
%NEW
In order to give the expressions for $\sigma_{xx}(\epsilon_i)$, $\sigma_{xv}(\epsilon_i)$ and $\sigma_{vv}(\epsilon_i)$ given in eq.~\eqref{sig} and~\eqref{sig2} we define the diffusion coefficient $D = {k_BT}/{\beta}$, the cyclic frequencies of the damped and the undamped harmonic potential $\omega = ({\beta^2}/{4m^2}-{\alpha}/{m})^{\frac{1}{2}}$ and $\omega_0 = ({\alpha}/{m})^{\frac{1}{2}}$ respectively and the characteristic time in the presence of frictional forces $\tau = {m}/{\beta}$ is. The $\sigma_{i,j}$ (here $i,j = \{x,v\}$) depend explicitly on time $\epsilon_i \in [t_{E_i},t]$. We have

\begin{align}\label{sigma_xx}
\sigma_{xx}(\epsilon_i)^2=&\frac{D}{4\omega^2\omega_0^2 \tau^3}\cdot \biggl(4\omega^2\tau^2 - \frac{1}{2}\left [\exp\left(-\epsilon_i \tau^{-1} + 2\omega \epsilon_i\right)(1 + 2\omega\tau) \right.   \\
& \left.+ \exp\left(-\epsilon_i \tau^{-1} - 2\omega\epsilon_i\right)(1 - 2\omega\tau)\right]+( 1 - 4\omega^2\tau^2)\exp(-\epsilon_i \tau^{-1}) \biggl) \nonumber
\end{align}
\begin{align}
\sigma_{xv}(\epsilon_i)^2 = &\frac{D}{4\omega^2 \tau^3}\cdot \biggl(4\omega^2\tau^2 - \frac{1}{2}\left [\exp\left(-\epsilon_i \tau^{-1} + 2\omega \epsilon_i\right)(1 - 2\omega\tau) \right.  \nonumber \\
& \left.+ \exp\left(-\epsilon_i \tau^{-1} - 2\omega\epsilon_i\right)(1 + 2\omega\tau)\right]+( 1 - 4\omega^2\tau^2)\exp\left(-\epsilon_i \tau^{-1}\right) \biggl) 
\end{align}
and
\begin{align}
\sigma_{vv}(\epsilon_i)^2 = \frac{D}{\omega^2\tau^2}\biggl(\exp\left(-\epsilon_i\tau^{-1} + 2\omega\epsilon_i\right)+\exp\left(-\epsilon_i\tau^{-1} - 2\omega\epsilon_i\right )- 2\exp\left(-\epsilon_i \tau^{-1}\right)\biggl).
\end{align}
Here, $\sigma_{xx}$, $\sigma_{xv}$ and $\sigma_{vv}$ are the elements of the variance-covariance matrix which fully characterizes a Gaussian distribution \cite{gillespie1996,norrelykke2011}.

\begin{acknowledgement}
This work was supported by the Deutsche
Forschungsgemeinschaft (DFG) within the collaborative
research center SFB 1027 and the research training group GRK 1276.
\end{acknowledgement}
\begin{table*}[hb]
\begin{center}
\begin{tabular}{|c|c|c|c|} 
\hline
 & \textbf{kinesin} & \textbf{dynein} & Ref.\\ \hline \hline
 $d$ & \multicolumn{2}{c|}{8 nm}  & \cite{Carter,toba2006}\\ \hline
 $N_\pm$ &  \multicolumn{2}{c|}{5} & \cite{welte1998} \\ \hline
 $L_0$&  \multicolumn{2}{c|}{110 nm} & \cite{kunwar2011} \\ \hline
 $v_{f}$ & \multicolumn{2}{c|}{$1000 \ \textnormal{nm/s} $} & \cite{Carter,toba2006}  \\ \hline
$v_{b}$ & \multicolumn{2}{c|}{$6 \ \textnormal{nm/s}$} & \cite{Carter,Gennerich}$^*$\\ \hline
$\alpha$ &  \multicolumn{2}{c|}{0.1 $\textnormal{pN/nm}$}& \cite{kunwar2011}$^*$\\ \hline
 $k_a$ & \multicolumn{2}{c|}{${5 \ }{\text{s}}^{-1}$}& \cite{mueller_k_l2008,leduc2004} \\ \hline
 $k_d^0$ & \multicolumn{2}{c|}{$1\ {\text{s}}^{-1}$} & \cite{kunwar2011}$^*$\ \\ \hline
  $f$ & \multicolumn{2}{c|}{1 pN} & \\ \hline
$F_S$ & 2.6 pN & 0.3-1.2 pN &\cite{Mallik2004,Shubeita} \\ \hline
  $k_\text{cat}^0$ &   \multicolumn{2}{c|}{$v_f\cdot d^{-1}$} & \cite{schnitzer_v_b2000} \\ \hline
  $k_\text{b}^0$ &   \multicolumn{2}{c|}{1.3 $\mu$M$^{-1}$s$^{-1}$}& \cite{schnitzer_v_b2000}\\ \hline
  $q_\text{cat}$ &  \multicolumn{2}{c|}{$6.2\cdot 10^{-3}$}& \cite{schnitzer_v_b2000} \\ \hline
  $ q_\text{b}$ & \multicolumn{2}{c|}{$4\cdot 10^{-2}$} & \cite{schnitzer_v_b2000}\\ \hline
$\Delta$ &4267.3 nm & $\max\left(\frac{2.9\cdot 10^5}{[ATP]^{0.3}}-28111.1,8534.6\right)$ nm & eq. (\ref{delta})\\ \hline  \hline
\multicolumn{4}{|c|}{Environment} \\ \hline \hline
$k_s$ & \multicolumn{2}{c|}{ $10^6$ s$^{-1}$ }& \ \\ \hline 
$\eta$ &  \multicolumn{2}{c|}{ 10 mPa$\cdot$s} & \cite{kulic2008}$^*$\\ \hline 
[ATP] &  \multicolumn{2}{c|}{ 0.5 mM} & \\ \hline 
$T$ &\multicolumn{2}{c|}{ 300 K } & \\ \hline 
$x_B$ &\multicolumn{2}{c|}{ 50 nm } & \\ \hline \hline
\multicolumn{4}{|c|}{Cargo} \\ \hline \hline
$R$ &  \multicolumn{2}{c|}{$1000$ nm }& \cite{thiam2013}$^*$ \\ \hline
$m$ &  \multicolumn{2}{c|}{$ 10^{-14}$ kg }&\\ \hline
\end{tabular}
\caption{The second and third column show the simulation parameters for kinesin and dynein, respectively. The fourth column gives the references providing experimental basis to these values. The $^*$ indicates that the experimental {values must be considered as orders of magnitude.}}
\label{paraset}
\end{center}
\end{table*}

\bibliographystyle{epj}
\bibliography{biblio.bib}

\begin{thebibliography}{32}

\bibitem{alberts2002}
B.~Alberts, A.~Johnson, J.~Lewis, M.~Raff, K.~Roberts, P.~Walter,
  \emph{{Molecular biology of the cell}}, 4th~edn. (Garland Science Taylor \&
  Francis Group, 2002), ISBN 0815332181

\bibitem{Carter}
N.J. Carter, R.A. Cross, Nature \textbf{435}(7040), 308 (2005)

\bibitem{vale2000}
R.D. Vale, R.A. Milligan, Science \textbf{288}(5463), 88 (2000)

\bibitem{welte1998}
M.A. Welte, S.P. Gross, M.~Postner, S.M. Block, E.F. Wieschaus, Cell
  \textbf{92}(4), 547 (1998)

\bibitem{soppina2009}
V.~Soppina, A.K. Rai, A.J. Ramaiya, P.~Barak, R.~Mallik, Proceedings of the
  National Academy of Sciences \textbf{106}(46), 19381 (2009)

\bibitem{hollenbeck_s2005}
P.J. Hollenbeck, W.M. Saxton, Journal of cell science \textbf{118}(23), 5411
  (2005)

\bibitem{welte2004}
M.A. Welte, Current Biology \textbf{14}(13), R525 (2004)

\bibitem{mueller_k_l2008}
M.J.I. M\"{u}ller, S.~Klumpp, R.~Lipowsky, Proceedings of the National Academy
  of Sciences \textbf{105}(12), 4609 (2008)

\bibitem{kulic2008}
I.M. Kuli{\'c}, A.E. Brown, H.~Kim, C.~Kural, B.~Blehm, P.R. Selvin, P.C.
  Nelson, V.I. Gelfand, Proceedings of the National Academy of Sciences
  \textbf{105}(29), 10011 (2008)

\bibitem{caspi_g_e2002}
A.~Caspi, R.~Granek, M.~Elbaum, Physical Review E \textbf{66}(1), 011916 (2002)

\bibitem{salman2002}
H.~Salman, Y.~Gil, R.~Granek, M.~Elbaum, Chemical physics \textbf{284}(1), 389
  (2002)

\bibitem{EPL14}
S.~Klein, C.~Appert-Rolland, L.~Santen, EPL \textbf{107}(1), 18004 (2014)

\bibitem{weiss2004}
M.~Weiss, M.~Elsner, F.~Kartberg, T.~Nilsson, Biophysical journal
  \textbf{87}(5), 3518 (2004)

\bibitem{reza2014}
M.R. Shaebani, Z.~Sadjadi, I.M. Sokolov, H.~Rieger, L.~Santen, submitted
  (2014)

\bibitem{korn2009}
C.B. Korn, S.~Klumpp, R.~Lipowsky, U.S. Schwarz, The Journal of chemical
  physics \textbf{131}(24), 245107 (2009)

\bibitem{kunwar2008}
A.~Kunwar, M.~Vershinin, J.~Xu, S.P. Gross, Current biology \textbf{18}(16),
  1173 (2008)

\bibitem{kunwar2011}
A.~Kunwar, S.K. Tripathy, J.~Xu, M.K. Mattson, P.~Anand, R.~Sigua,
  M.~Vershinin, R.J. McKenney, C.C. Yu, A.~Mogilner et~al., Proceedings of the
  National Academy of Sciences \textbf{108}(47), 18960 (2011)

\bibitem{schnitzer_v_b2000}
M.J. Schnitzer, K.~Visscher, S.M. Block, Nat Cell Biol \textbf{2}(10), 718
  (2000)

\bibitem{toba2006}
S.~Toba, T.M. Watanabe, L.~Yamaguchi-Okimoto, Y.Y. Toyoshima, H.~Higuchi,
  Proceedings of the National Academy of Sciences \textbf{103}(15), 5741 (2006)

\bibitem{mallik_g2004}
R.~Mallik, S.P. Gross, Current Biology \textbf{14}(22), R971  (2004), ISSN
  0960-9822

\bibitem{gillespie1996}
D.T. Gillespie, Physical review E \textbf{54}(2), 2084 (1996)

\bibitem{norrelykke2011}
S.F. N{\o}rrelykke, H.~Flyvbjerg, Physical Review E \textbf{83}(4), 041103
  (2011)

\bibitem{TGF13}
S.~Klein, C.~Appert-Rolland, L.~Santen, to appear in TGF13  (2014)

\bibitem{gillespie1978}
D.T. Gillespie, Journal of Computational Physics \textbf{28}(3), 395 (1978)

\bibitem{luby1999}
K.~Luby-Phelps, International review of cytology \textbf{192}, 189 (1999)

\bibitem{redner2001}
S.~Redner, \emph{A guide to first-passage processes} (Cambridge University
  Press, 2001)

\bibitem{mckenney2014}
R.J. McKenney, W.~Huynh, M.E. Tanenbaum, G.~Bhabha, R.D. Vale, Science p.
  1254198 (2014)

\bibitem{Gennerich}
A.~Gennerich, A.P. Carter, S.L. Reck-Peterson, R.D. Vale, Cell \textbf{131}(5),
  952 (2007)

\bibitem{leduc2004}
C.~Leduc, O.~Camp{\`a}s, K.B. Zeldovich, A.~Roux, P.~Jolimaitre,
  L.~Bourel-Bonnet, B.~Goud, J.F. Joanny, P.~Bassereau, J.~Prost, Proceedings
  of the National Academy of Sciences of the United States of America
  \textbf{101}(49), 17096 (2004)

\bibitem{Mallik2004}
R.~Mallik, B.C. Carter, S.A. Lex, S.J. King, S.P. Gross, Nature
  \textbf{427}(6975), 649 (2004)

\bibitem{Shubeita}
G.T. Shubeita, S.L. Tran, J.~Xu, M.~Vershinin, S.~Cermelli, S.L. Cotton, M.A.
  Welte, S.P. Gross, Cell \textbf{135}(6), 1098  (2008)

\bibitem{thiam2013}
A.R. Thiam, R.V. Farese~Jr, T.C. Walther, Nature Reviews Molecular Cell Biology
  \textbf{14}(12), 775 (2013)

\end{thebibliography}

\end{document}